\documentclass[12pt]{revtex4}
\usepackage{graphicx}
\usepackage{amssymb}

\newcommand{\s}{\mathcal{S}}

\newcommand{\ga}{g_{\chi}}
\newcommand{\gb}{g_{\mathcal{S}}}
\newcommand{\gc}{g_{\chi\mathcal{S}}}

\begin{document}
\topmargin 0.1in
\textheight 8.6in
\textwidth 6.6in
\thispagestyle{empty}
\begin{flushright} 
UCRHEP-T443\\ 
November 2007\
\end{flushright}
\vskip 0.5in

\title{Multipartite Dark Matter}

\author{Qing-Hong Cao}
\affiliation{Department of Physics and Astronomy, University of California, 
Riverside, CA 92521}
\author{Ernest Ma}
\affiliation{Department of Physics and Astronomy, University of California, 
Riverside, CA 92521}
\author{Jos\'{e} Wudka}
\affiliation{Department of Physics and Astronomy, University of California, 
Riverside, CA 92521}
\author{C.-P. Yuan}
\affiliation{Department of Physics and Astronomy, Michigan State University, 
East Lansing, MI 48824}

\begin{abstract}
Dark matter (comprising a quarter of the Universe) is usually assumed 
to be due to \emph{one and only one} weakly interacting particle which 
is neutral and absolutely stable.  We consider the possibility that there 
are several coexisting dark-matter particles, and explore in some detail 
the generic case where there are \emph{two}. We discuss how the second 
dark-matter particle may relax the severe constraints on the parameter 
space of the Minimal Supersymmetric Standard Model (MSSM), as well as 
other verifiable predictions in both direct and indirect search experiments. 
\end{abstract}
\maketitle

\newpage

Dark matter (DM) is at the heart of any study regarding the interface
between particle physics, astrophysics, and cosmology. Its relic abundance
has now been measured with precision. Combining the results of the
WMAP Collaboration and the Sloan Digital Sky Survey, 
$\Omega_{CDM}h^{2}=0.110\pm0.013\,\,(2\sigma)$\ \cite{WMAP},
where $\Omega_{CDM}$ is the DM energy density normalized by the critical
density of the Universe and $h=0.71\pm0.05\,\,(2\sigma)$ is the scaled
Hubble parameter. Many dark-matter candidates have been suggested
in various models beyond the Standard Model (SM) of particle physics,
but a nearly universal implicit assumption is that one and only one
such candidate (1DM) is needed and its properties are constrained
accordingly. This is of course not a fundamental principle and the
possibility of \emph{multipartite dark matter} should not be ignored.
In this Letter we study \emph{its impact on the conventional picture
of 1DM physics}, such as that of supersymmetry, using a simple generic
scenario of two dark-matter candidates (2DM), one a fermion singlet
(neutralino) and the other a scalar singlet. Our conclusions are broadly
applicable to any 2DM model.

\textbf{Model} ~The simplest way to have at least two DM candidates
is to append the SM with the exactly conserved discrete symmetry 
$\mathbb{Z}_{2}\times\mathbb{Z}_{2}^{\prime}$.
As pointed out in Ref.\ \cite{Boehm}, this may be realized naturally in the 
framework of $N=2$ supersymmetry. Alternatively, if the SM is extended to 
include an exactly conserved $\mathbb{Z}_{2}$ symmetry without supersymmetry, 
then the supersymmetric version of this extension will have 
$\mathbb{Z}_{2}\times\mathbb{Z}_{2}^{\prime}$, 
as in Refs.\ \cite{Ma} and \cite{hln07}. To explore generically 
the impact of such a scenario,
we first observe that the details of the specific model are mostly
irrelevant, as far as the relic abundance, and the direct and indirect
detection of dark matter are concerned, except for the masses of the
two DM candidates and their interactions with the SM particles and
with each other. This is because the relevant processes are 
either elastic scattering at almost zero momentum transfer or 
annihilation at rest.

Specifically we add two new fields which are singlets under the SM
gauge group: a new fermion $\chi$ and a new scalar $\s$. Under 
$\mathbb{Z}_{2}\times\mathbb{Z}_{2}^{\prime}$,
$\chi\sim(-,+)$ and $\s\sim(+,-)$, whereas all SM particles are
$(+,+)$. This means that $\left\langle \s\right\rangle =0$ is required.
In a complete theory such as that of Ref.\ \cite{Ma}, there may
also be $(-,-)$ particles. For simplicity we assume that all such
particles are heavy enough to decay into $\chi$ and $\s$. If not,
we would then have to consider three coexisting DM candidates.

The Lagrangian of our generic 2DM model is given by 
\begin{equation}
\mathcal{L}=\mathcal{L}_{SM}+\mathcal{L}_{DM}^{\chi}+\mathcal{L}_{DM}^{\s}
+\mathcal{L}_{int},
\end{equation}
where $\mathcal{L}_{SM}$ denotes the usual SM Lagrangian, and 
\begin{eqnarray}
&  & \mathcal{L}_{DM}^{\chi}=i\bar{\chi}\not\!\partial\chi-m_{1}\bar{\chi}\chi,
\nonumber \\
&  & \mathcal{L}_{DM}^{\s}=\frac{1}{2}\partial_{\mu}\s\partial^{\mu}\s
+\frac{1}{2}m_{2}^{2}\s^{2}+\frac{1}{4}\lambda_{1}\s^{4}, \nonumber \\
&  & \mathcal{L}_{int}=\frac{1}{2}\lambda_{2}H^{\dagger}H\s\s
+\frac{\lambda_{3}}{\Lambda}H^{\dagger}H\bar{\chi}\chi
+\frac{\lambda_{4}}{2\Lambda}\bar{\chi}\chi\s\s,
\end{eqnarray}
where $H$ is the SM Higgs doublet. 
After the electroweak symmetry is spontaneously broken, 
$H=(v+h)/\sqrt{2}$ with $v=246$ GeV
and the masses of $\chi$ and $\s$ are given by 
$m_{\chi}=m_{1}-\lambda_{3}v^{2}/2\Lambda$ 
and $m_{\s}^{2}=m_{2}^{2}+\lambda_{2}v^{2}/2$.
The various effective interaction terms, relevant to our discussion, are
\begin{eqnarray}
\mathcal{L}_{h\chi\chi}=\ga h\bar{\chi}\chi, &  & \mathcal{L}_{hh\chi\chi}
=\frac{\ga}{2v}hh\bar{\chi}\chi,\nonumber \\
\mathcal{L}_{h\s\s}=\frac{1}{2}\gb v\, h\s\s, &  & \mathcal{L}_{hh\s\s}
=\frac{1}{4}\gb\, hh\s\s,\nonumber \\
\mathcal{L}_{\chi\chi\s\s}=\frac{\gc}{v}\bar{\chi}\chi\s\s,
\end{eqnarray}
where we have introduced the dimensionless couplings $\ga=\lambda_{3}v/\Lambda$,
$\gb=\lambda_{2}$, and $\gc=\lambda_{4}v/2\Lambda$.  Note that this is not 
meant to be an effective theory in powers of $1/\Lambda$ for all processes. 
It is applicable only to DM-nucleus elastic scattering 
(with almost zero momentum transfer) and DM annihilation at rest.

\begin{figure}[htb]
\includegraphics[clip,scale=0.6]{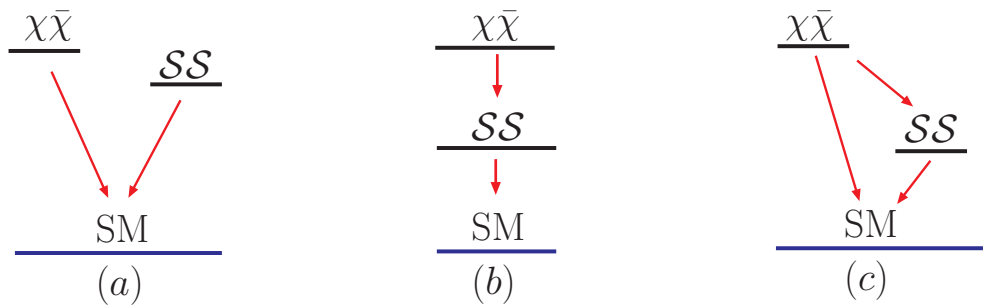}
\caption{Possible annihiliation scenarios in the 2DM model where the (red) arrow
line denotes the DM annihilation. \label{fig:scheme}}
\end{figure}

As an example of how the effective couplings of Eq.~(3) may be generated
in a complete model, let us consider Ref.\ \cite{Ma}, where a second pair
of scalar superfields $(\eta_{1}^{0},\eta_{1}^{-})$ and 
$(\eta_{2}^{+},\eta_{2}^{0})$
are added, which are odd under a new $\mathbb{Z}_{2}$, whereas the
usual $(\phi_{1}^{0},\phi_{1}^{-})$ and $(\phi_{2}^{+},\phi_{2}^{0})$
of the MSSM are even. Together with the conventional $R$ parity,
we then have an exactly conserved $\mathbb{Z}_{2}\times\mathbb{Z}_{2}^{\prime}$
symmetry. Consider now the interaction 
\begin{equation}
\frac{1}{2}g_{Y}\widetilde{B}(\widetilde{\eta}_{1}^{0}\eta_{1}^{0}
-\widetilde{\eta}_{2}^{0}\eta_{2}^{0})
+\frac{1}{2}g_{Y}\widetilde{B}(\widetilde{\phi}_{1}^{0}\phi_{1}^{0}
-\widetilde{\phi}_{2}^{0}\phi_{2}^{0}),
\end{equation}
where $\widetilde{B}$ is the $U(1)_{Y}$ gaugino. We may thus identify
the $\chi$ of our generic model with $\widetilde{B}$, and $\s$($h$)
with a linear combination of the real parts of $\eta_{1,2}^{0}$($\phi_{1,2}^{0}$).
The effective $\bar{\chi}\chi\s\s$ and $hh\bar{\chi}\chi$ interactions are then
generated from the exchange of the $\widetilde{\eta}$ and $\widetilde{\phi}$
higgsinos, respectively. Assuming the masses of these higgisinos to be comparable to
$m_{\chi}$ and $m_{\s}$, the effective couplings $\ga$ and $\gc$
are not necessarily very much suppressed. This allows us to consider
three characteristic scenarios, as depicted in Fig.\ \ref{fig:scheme}.
For definiteness, we consider $m_{\chi}>m_{\s}$ in this analysis,
but our conclusions are mostly the same if we switch them around.

Scenario A [Fig.\ 1(a)]: $\ga,\gb\neq0$ but $\gc=0$; both $\chi$ and $\s$ 
can annihilate into SM particles but they do not interact with each other. 
Scenario B [Fig.\ 1(b)]: $\ga=0$ but $\gb,\gc\neq0$; $\chi$ can only 
annihilate into $\s$, after which $\s$ will annihilate into SM particles. 
All of $\chi$'s information is hidden behind $\s$, hence it will be 
\emph{superdark} and cannot be observed directly.  It may be revealed 
nevertheless if apparent discrepancies occur among different experiments.
Scenario C [Fig.\ 1(c)]: $\ga,\gb,\gc\neq0$; $\chi$ can annihilate
into both $\s$ and SM particles, after which $\s$ will annihilate into SM
particles.  The special case of $\gc\gg\ga,\gb$ is of particular
interest. Here $\chi\bar{\chi}$ will annihilate predominantly into $\s\s$,
resulting in a much smaller $\chi$ relic abundance, thereby relaxing
the constraints on its parameter space, which may be identified
with that of the MSSM. Scenarios A and B are of course just two special 
limits of C, but they have qualitatively different predictions on the
direct and indirect search experiments of dark matter, as shown below.

\textbf{Observational Constraints} ~If two DM candidates coexist,
the usual observational constraints also apply, but with modifications.

\emph{(i) Relic abundance}: Since both DM candidates contribute to
the relic abundance, they must add up to account for the current observation:
\begin{equation}
\Omega_{\chi}h^{2}+\Omega_{\s}h^{2}=\Omega_{CDM}h^{2}=0.110\pm0.013.
\label{eq:relic}
\end{equation}
It is well-known that the relic density of each DM species is approximately
given by $\Omega_{i}h^{2}\approx(0.1\,{\rm pb})/
\left\langle \sigma v\right\rangle _{i}$,
where $\left\langle \sigma v\right\rangle _{i}$ is the thermally
averaged product of its annihilation cross section with its velocity.
Using Eq.\ (\ref{eq:relic}), we then obtain 
\begin{equation}
\frac{\left\langle \sigma v\right\rangle _{\chi}
\left\langle \sigma v\right\rangle _{\s}}
{\left\langle \sigma v\right\rangle _{\chi}
+\left\langle \sigma v\right\rangle _{\s}}\equiv
\left\langle \sigma v\right\rangle _{0}\sim{\rm pb}.
\label{eq:relic-xsec}
\end{equation}

\emph{(ii) Halo density profile:} For simplicity, we assume the two
DM candidates to have the same density profile and use that given
by Navarro, Frenck and White (NFW)\ \cite{NFW} in our analysis.
(It is of course straightforward to extend our results to other density
profiles.) In the 2DM model, the dark-matter mass density profile
of the galactic halo is thus given by 
\begin{equation}
\rho(r)=\frac{\epsilon_{\chi}\rho_{0}}{\left(r/r_c\right)\left(1+r/r_c\right)^{2}}
+\frac{\epsilon_{\s}\rho_{0}}{\left(r/r_c\right)\left(1+r/r_c\right)^{2}},
\label{eq:profile}
\end{equation}
where $r_c=20.0\,{\rm kpc}$ and $\rho_{0}$ is adjusted to reproduce
the local halo density at the Earth position. Here, $\epsilon_{i}$
represents the \emph{fraction} of the mass density of the $i$th dark
matter in our local dark-matter halo as well as in the Universe, i.e.
\begin{equation}
\epsilon_{i}=\frac{\rho_{i}}{\rho_{0}}\simeq\frac{\Omega_{i}h^{2}}
{\Omega_{CDM}h^{2}},\label{eq:fraction}
\end{equation}
where $\rho_{i}$ is the local density of the $i$th DM and 
$\sum_{i}\epsilon_{i}=1$. For our 2DM model, we obtain 
\begin{equation}
\epsilon_{\chi}=\frac{\left\langle \sigma v\right\rangle _{0}}
{\left\langle \sigma v\right\rangle _{\chi}},\quad\epsilon_{\s}
=\frac{\left\langle \sigma v\right\rangle _{0}}
{\left\langle \sigma v\right\rangle _{\s}}.\label{eq:fraction-xsec}
\end{equation}

\emph{(iii) Direct search}: Assuming that DM is the dominant component
of the halo of our galaxy, it is expected that a certain number of
these weakly interacting massive particles (WIMPs) will cross the
Earth at a reasonable rate and be detected by measuring the energy
deposited in a low-background detector through the scattering of a
WIMP with a nucleus of the detector. So far most experimental limits
of this direct detection are given in terms of the cross section per
nucleon under the 1DM hypothesis. The event rate per unit time per
nucleon is given by 
\begin{equation}
R\approx\sum_{i}n_{i}\left\langle \sigma\right\rangle _{i}
=\sum_{i}\frac{\rho_{i}}{m_{i}}\left\langle \sigma\right\rangle _{i},
\label{eq:directrate}
\end{equation}
where $n_{i}$ is the local number density of the $i$th DM and 
$\left\langle \sigma\right\rangle _{i}$
is the $i$th DM-nucleon elastic scattering cross section which
is averaged over the relative DM velocity with respect to the detector.
The measured experimental rate in the 1DM case is given by 
$R_{{\rm exp}}\approx\rho_{0}\sigma_{0}/m_{0}$
where $\sigma_{0}$ denotes the {}``zero-momentum-transfer'' cross
section of DM-nucleon scattering and $m_{0}$ is the DM mass. The
current direct-search limit implies $R<R_{{\rm exp}}$, i.e. 
\begin{equation}
\frac{\epsilon_{\chi}}{m_{\chi}}\sigma_{\chi\mathcal{N}}
+\frac{\epsilon_{\s}}{m_{\s}}\sigma_{\s\mathcal{N}}<\frac{\sigma_{0}}{m_{0}},
\label{eq:directxsec}
\end{equation}
where $\sigma_{\chi\mathcal{N}}$($\sigma_{\s\mathcal{N}}$) denotes
the scattering cross section of $\chi$($\s$) with a nucleon $\mathcal{N}$.
Although the experimental sensitivities and limits are often described
in terms of the dark-matter elastic scattering with a single nucleon,
one should keep in mind that nuclear form factors may need to be taken
into account.

In Scenario B, there is no scattering of $\chi$ with the nucleon, 
hence the limit in Eq.\ (\ref{eq:directxsec}) becomes 
$\sigma_{\s\mathcal{N}}<\sigma_{0}/\epsilon_{\s}$,
i.e. bounds from direct detection become weaker in this case.  On the other 
hand, if dark matter is observed in direct-detection experiments,
the DM-neucleon cross section may be understimated by a factor of 
$1/\epsilon_{\s}$. 

\emph{(iv) Indirect gamma-ray search:} The relic dark matter may collect
and become gravitationally bound to the center of the galaxy, the
center of the Sun and the center of the Earth. If this happens, then
a variety of \emph{indirect} dark-matter detection opportunities arise.
In particular, the measurement of secondary particles coming from
dark-matter annihilation in the halo of the galaxy will help to decipher
the nature of dark matter. Efforts to detect the annihilation products
of dark-matter particles in the form of gamma rays, antimatter and
neutrinos are collectively known as indirect detection. Of these,
the observation through gamma rays is the simplest and most robust.
The diffusion gamma-ray spectrum is given by 
\begin{equation}
\frac{d\Phi}{dE_{\gamma}}=\epsilon_{\chi}^{2}\frac{d\Phi_{\chi}}
{dE_{\gamma}}+\epsilon_{\s}^{2}\frac{d\Phi_{\s}}{dE_{\gamma}},
\label{eq:gammaray}
\end{equation}
where $d\Phi_{i}/dE_{\gamma}$ ($i=\chi,$ $\s$) is the differential
gamma-ray flux along a direction that forms an angle $\psi$ with
respect to the direction of the galactic center: 
\begin{equation}
\frac{d\Phi_{i}}{dE_{\gamma}}=\frac{dN_{\gamma}}{dE_{\gamma}}
\left\langle \sigma v\right\rangle _{i}\frac{1}{4\pi m_{i}^{2}}
\int_{\psi}\left[\frac{\rho_{0}}{\left(r/r_{c}\right)
\left(1+r/r_{c}\right)^{2}}\right]^{2}dl.
\end{equation}
The integral is performed along the line of sight. All annihilation 
channels of the $i$th DM are summed, and $dN_{\gamma}/dE_{\gamma}$ is the 
differential gamma spectrum per annihilation coming from the decay 
of annihilation products.

Consider the special case of $m_{\chi}=m_{\s}=m_{0}$. After some
simple algebra, one can show that \begin{equation}
\frac{d\Phi}{dE_{\gamma}}\simeq\frac{dN_{\gamma}}{dE_{\gamma}}
\left\langle \sigma v\right\rangle _{0}\frac{1}{4\pi m_{0}^{2}}
\int_{\psi}\left[\frac{\rho_{0}}
{\left(r/r_c\right)\left(1+r/r_c\right)^{2}}\right]^{2}dl,\end{equation}
where we have used the fact that $dN_{\gamma}/dE_{\gamma}$ is almost
the same for most of the final states. The integrated flux of the
2DM model is of the same order as that of the 1DM model.

\emph{(v) Collider search}: Since 
$\Omega h^{2}\propto1/\left\langle \sigma v\right\rangle $,
the requirement of the correct relic density ($\Omega_{\rm CDM}h^{2}\sim0.1$)
implies that DM annihilation was efficient in the early Universe.
It also suggests efficient annihilation now, implying large indirect
detection rates, as well as efficient scattering now, implying large
direct detection rates. The sum rule, cf. Eq.\ (\ref{eq:relic}), means
that the DM annihilation of each individual candidate has to be more
efficient than that of the 1DM case. Hence larger cross sections of 
DM production are expected at the collider. The smaller
the fraction $\epsilon_{i}$, the easier is the detection. In our
simplistic case where the two DM candidates interact only with the
SM Higgs boson, the vector-boson-fusion process $qq\to qqVV\to qqh$,
with the subsequent decay $h\to\chi\bar{\chi}/\s\s$, provides the most 
promising collider signature of the model\ \cite{DSB} 
when $m_h>m_{\chi/\s}$. 

\begin{figure}[t]
\includegraphics[clip,scale=0.8]{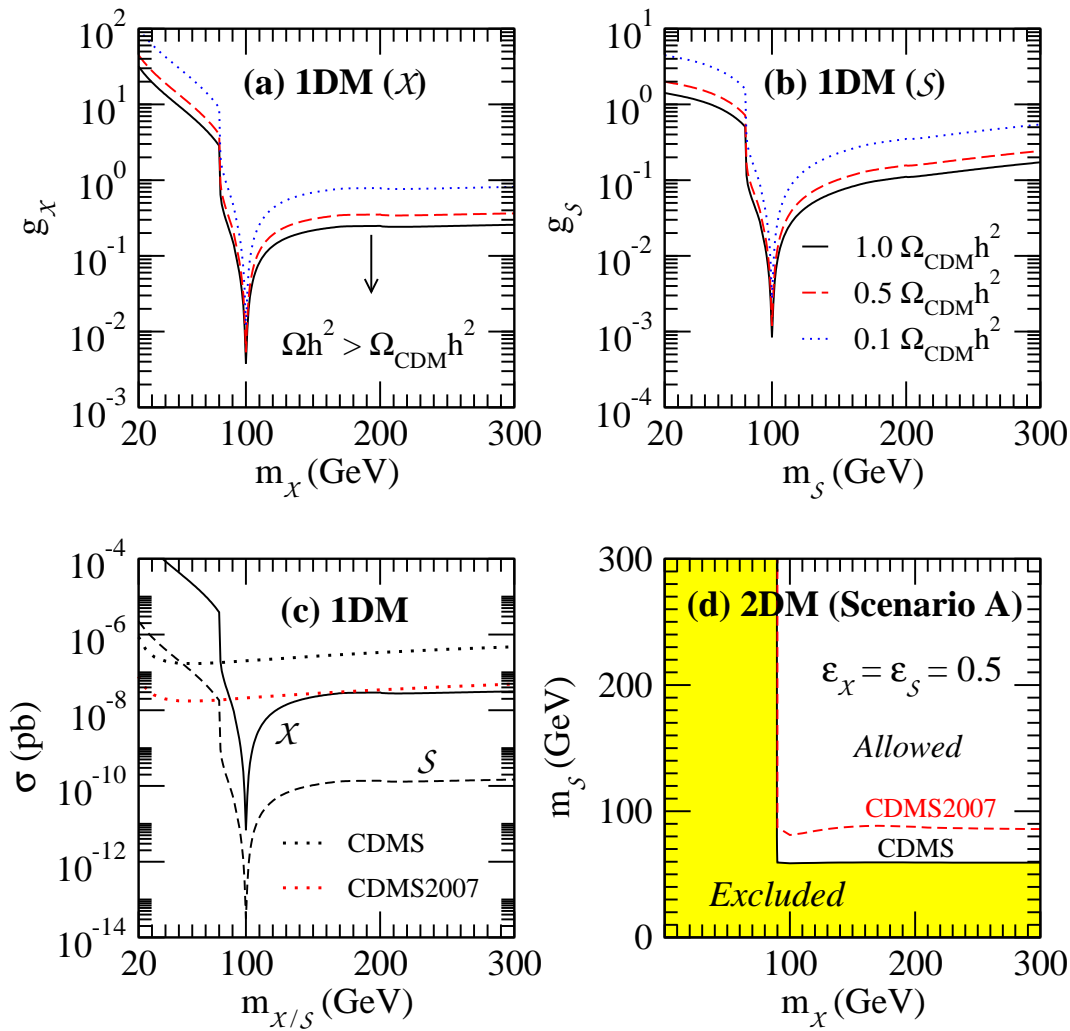}
\caption{(a) and (b) show the correlations between the coupling and the mass
of a single DM candidate as determined by the WMAP data: (a) for $\chi$
and (b) for $\s$. (c) shows the spin-independent cross sections of
DM-nucleon scattering in the 1DM model together with the CDMS limit
and future projected sensitivities of CDMS2007. 
We choose $m_{h}=200~\rm{GeV}$ throughout in this work. (d) shows
the allowed ($m_{\chi},\, m_{\s}$) parameter space of Scenario A in the 2DM
model. \label{fig:fig1}}
\end{figure}

\textbf{2DM Implications} ~We first study the cosmological implications
of either $\chi$ or $\s$ as the sole source of dark matter. 
In Fig.\ \ref{fig:fig1}(a)
and (b) we present the correlations between the effective coupling
and the DM mass\ \cite{1DM}, which is derived from WMAP data. The
black-solid (red-dashed, blue-dotted) curve denotes 
$\Omega_{i}h^{2}\simeq0.1\,(0.05,\,0.01)$,
respectively. In the region below the black-solid curve
the dark matter is overproduced.
Fig.\ \ref{fig:fig1}(c) shows the spin-independent
cross section of DM-nucleon scattering for $\chi$ (black-solid) and
$\s$ (black-dashed). Current CDMS limit and projected sensitivity
of CDMS2007\ \cite{CDMS} are also plotted. Using Eq.\ (\ref{eq:directxsec}),
we then derive a realistic bound on the 2DM model . For a large range
of the DM mass, $\sigma_{0}/m_{i}$ is almost a constant, e.g. 
$\sigma_{0}/m_{i}\simeq2\times10^{-9}\,{\rm pb}/{\rm GeV}
\,\left(2\times10^{-10}\,{\rm pb}/{\rm GeV}\right)$
for the current CDMS data (projected CDMS2007 sensitivity). 
In Fig.\ \ref{fig:fig1}(d) we present the allowed parameter space
of Scenario A in the 2DM model in the 
($m_{\s},\, m_{\chi}$) plane for $\epsilon_{\chi}=\epsilon_{\s}=0.5$.
In Scenario B, the limits only depend on
$\s$ and the bounds become weaker.

\begin{figure}[t]
\includegraphics[clip,scale=0.9]{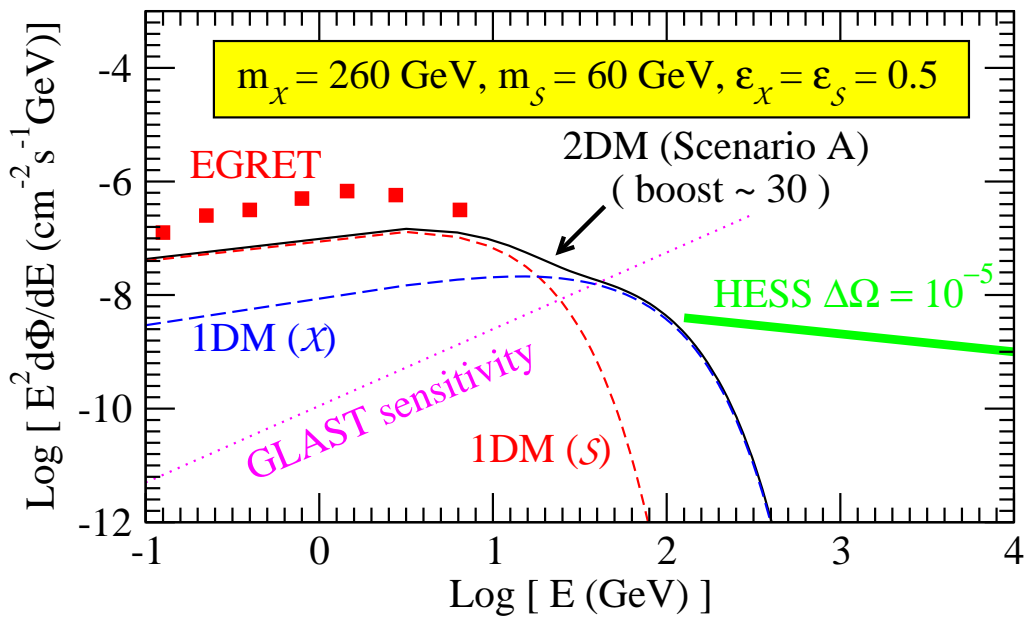}
\caption{Predicted gamma-ray spectra in the 2DM model for 
$\epsilon_{\chi}=\epsilon_{\s}=0.5$.
The predicted gamma flux is from a $\Delta\Omega=10^{-3}\, srad$ region
around the direction of the galactic center, assuming the NFW halo
profile (with a boost factor as indicated in the figure). For comparison
we also show the scaled gamma-ray distribution in the 1DM case. 
EGRET and HESS observations are also shown here for comparison. 
\label{fig:gamma-ray}}

\end{figure}

A promising way to tell if there are two coexisting DM candidates,
assuming that they are very different in mass, is through indirect
gamma-ray observations. The overlap of the two distributions might
change the line shape of the gamma-ray distribution which is distinguishable
from that of the 1DM case. In Fig.\ \ref{fig:gamma-ray} this
is illustated by showing the predicted fluxes from a 
$\Delta\Omega=10^{-3}\,{\rm srad}$
region around the direction of the galactic center together with the
existing EGRET\ \cite{EGRET} and HESS\ \cite{HESS} observations
in the same sky direction. We adopt the NFW density profile for the
DM in our galaxy ($\bar{J}\times\Delta\Omega\sim1$ for 
$\Delta\Omega=10^{-3}\,{\rm srad}$)
and allow the flux to be scaled by a {}``boost factor''. For demonstration
we choose $\epsilon_{\chi}=\epsilon_{\s}=0.5$, $m_{\chi}=260\,{\rm GeV}$,
and $m_{\s}=60\,{\rm GeV}$. Clearly, the resulting gamma-ray flux distribution
from the overlap of 2DM distribution is significantly different from that of 
the 1DM model, which can be probed by the GLAST experiment\ \cite{GLAST}. 
The gamma-ray spectra can also be used to distinguish Scenario A from B 
of the 2DM model. Unfortunately, it is difficult to observe a shape change if 
$m_\chi - m_\s$ is small.  On the other hand, they may be discriminated at 
the Large Hadron Collider (LHC) because in Scenario A, both $\chi$ and 
$\s$ are produced; whereas in Scenario B, only $\s$ is.  A discrepancy 
between relic abundance and LHC production may reveal Scenario B.

\begin{figure}[t]
\includegraphics[clip,scale=0.9]{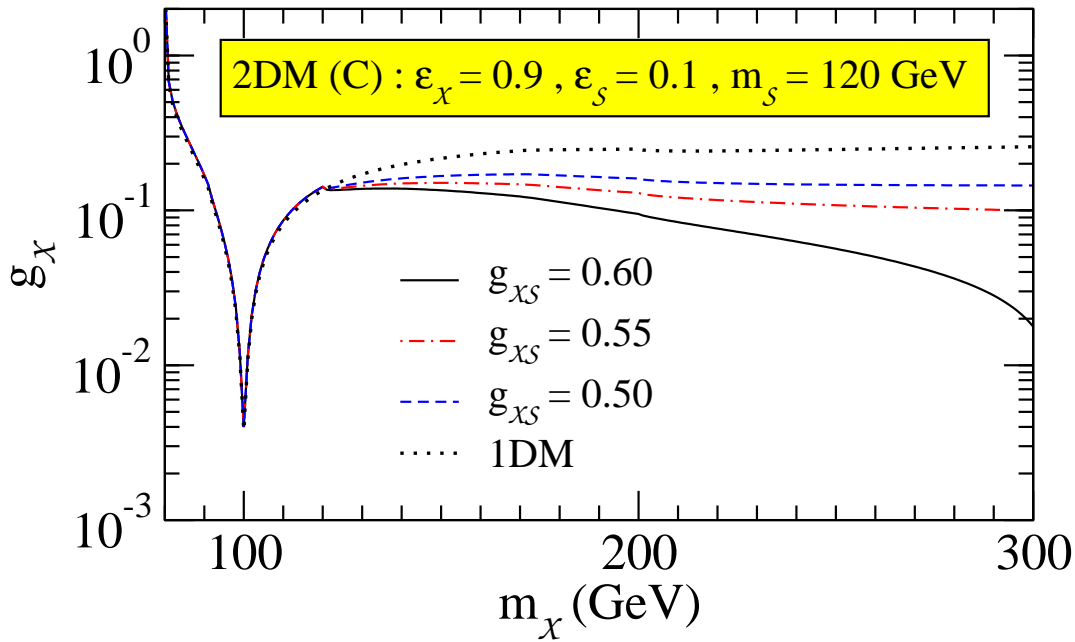}
\caption{Relation between $g_{\chi}$ and $m_{\chi}$ when the 
$\chi\bar{\chi}\to\s\s$ annihilation mode is open. 
For illustration, we choose 
$\epsilon_{\chi}(\epsilon_{\s})=0.9(0.1)$
and $m_{\s}=120\,{\rm GeV}$. \label{fig:d}}
\end{figure}

Consider the special case ($\gc\gg\ga,\,\gb$) of Scenario C, 
where the new annihilation channel $\chi\bar{\chi}\to\s\s$ opens.
This case is very interesting because it has a crucial impact on the
conventional supersymmetric DM model. For example, the lightest neutralino
is a well-motivated dark-matter candidate, but its relic abundance
is typically too large, or equivalently, its annihilation rate is
too small. The WMAP data thus impose very tight constraints on the
parameter space of the MSSM. But those constraints can be relaxed
if there exists an additional DM candidate which opens up a new annihilation
channel for the neutralino. For illustration, we choose $\gc=(0.5,0.55,0.6)$
and $m_{\s}=120\,{\rm GeV}$ with $\epsilon_{\chi}(\epsilon_{\s})=0.9(0.1)$
in the 2DM model. Using Fig.\ \ref{fig:fig1}(b), we then fix $\gb=0.114$.
The (black) dotted curve in Fig.\ \ref{fig:d} denotes
$\Omega_{\chi}h^{2}=0.1$ in the 1DM model and the region below it
will exceed the relic abundance.  After including the new annihilaton
channel $\chi\bar{\chi}\to\s\s$, more of the parameter space is reclaimed. 
Increasing $\gc$ will open up even more parameter space.

\textbf{Conclusion} ~In this Letter we presented a simple generic
model of two coexisting dark-matter candidates. We discussed its three
characteristic annihilation scenarios and its impact on the 
observational constraints of dark matter. We note that the
cosmic gamma-ray observation is a good probe for confirming the 2DM
model. We also demonstrate that with a second dark-matter candidate,
the usual severe constraints on the parameter space of the MSSM can be
relaxed. More detailed studies of this new idea of multipartite 
dark matter are forthcoming.

\emph{Acknowledgements} This work is supported in part by the U.~S.~Department
of Energy under Grant No.~DE-FG03-94ER40837 and the U.\ S.\ National
Science Foundation under award PHY-0555545.

\end{document}